\documentclass[twocolumn]{aastex7}
\usepackage{{booktabs}}
\usepackage{romannum}
\usepackage{float}
\usepackage{hyperref}
\usepackage{soul}
\usepackage{xfrac}
\usepackage{acronym}
\usepackage{todonotes}
\setulcolor{red}
\setstcolor{red}
\todostyle{Bibhuti}{author=\textbf{Bibhuti}, color=orange!20!white, shadow, inline}

\newcommand{\bkjq}[1]{\textcolor{red}{\bf (#1)}}

\def\fill{\bkjq{Fill}}

\newcommand{\ca}{\mbox{Ca\,{\sc ii}~K\,}}

\graphicspath{ {./figures/} }
\renewcommand{\textbf}[1]{\textcolor{magenta}{{\bf #1}}}

\begin{document}

\acrodef{sse}[SSE]{standard statistical error}
\def\sse{\ac{sse}}

\acrodef{cc}[CC]{co4rrelation coefficient}
\def\cc{\ac{cc}}

\acrodef{mdi}[MDI]{Michelson Doppler Imager}
\def\mdi{\ac{mdi}}

\def\cak{Ca {\sc ii} K}
\def\ha{H-$\alpha$}

\title{Machine Learning Based Identification of Solar Disk and Plages in Kodaikanal Solar Observatory Historical Suncharts}

\shorttitle{Plage Detection using Machine Learning}
\shortauthors{D. K. Mishra et al.}

\correspondingauthor{Dipankar Banerjee}
\email{dipu@iiap.res.in}

\author[0009-0003-1377-0653]{Dibya Kirti Mishra}
\affiliation{Aryabhatta Research Institute of Observational Sciences, Nainital-263002, Uttarakhand, India}
\affiliation{Department of Applied Physics, Mahatma Jyotiba Phule Rohilkhand University, Bareilly-243006, Uttar Pradesh, India}
\email[show]{dibyakirtimishra@aries.res.in}

\author[0000-0002-5014-7022]{Subhamoy Chatterjee}
\affiliation{Southwest Research Institute, Boulder, CO 80302, USA}
\email{subhamoy.chatterjee@swri.org}

\author[0000-0003-3191-4625]{Bibhuti Kumar Jha}
\affiliation{Southwest Research Institute, Boulder, CO 80302, USA}
\email{bibhuti.jha@swri.org}

\author[0000-0002-5829-2697]{Hemapriya Raju}
\affiliation{Tata Institute of Fundamental Research, Colaba, Mumbai 400005, Maharashtra}
\email{hemapriya.raju@tifr.res.in}

\author[0000-0003-2476-1536]{Aditya Priyadarshi}
\affiliation{Curtin University, Karawara, Western Australia, Australia}
\email{rbadityapriyadarshi@gmail.com}

\author[0000-0002-8163-3322]{Manjunath Hegde}
\affiliation{Indian Institute of Astrophysics, Koramangala, Bangalore 560034, India}
\email{manjunath.n.hegde@gmail.com}


\author[0009-0008-5834-4590]{Srinjana Routh}
\affiliation{Aryabhatta Research Institute of Observational Sciences, Nainital-263002, Uttarakhand, India}
\affiliation{Department of Applied Physics, Mahatma Jyotiba Phule Rohilkhand University, Bareilly-243006, Uttar Pradesh, India}
\email{srinjana.routh@gmail.com}

\author[0000-0003-4653-6823]{Dipankar Banerjee}
\affiliation{Indian Institute of Space Science and Technology, Thiruvananthapuram, 695 547 Kerala, India}
\affiliation{Indian Institute of Astrophysics, Koramangala, Bangalore 560034, India}
\affiliation{Center of Excellence in Space Sciences India, IISER Kolkata, Mohanpur 741246, West Bengal, India}
\email{dipu@iiap.res.in}

\author{M. Saleem Khan}
\affiliation{Department of Applied Physics, Mahatma Jyotiba Phule Rohilkhand University, Bareilly-243006, Uttar Pradesh, India}
\email{saleem.hepru@gmail.com}

\begin{abstract}
Kodaikanal Solar Observatory (KoSO) is one of the oldest solar observatories, possessing an archive of multi-wavelength solar observations, including white light, \ca, and \ha~images spanning over a century. In addition to these observations, KoSO has preserved hand-drawn suncharts (1904\,--\,2022), on which various solar features such as sunspots, plages, filaments, and prominences are marked on the Stonyhurst grid with distinct colour coding. In this study, we present the first comprehensive result that includes the entire data set from these suncharts using a supervised Machine Learning model called ``Convolutional Neural Networks (CNNs)", firstly to identify the solar disks from the charts (1909\,--\,2007), secondly to identify of the plages, spanning 9 solar cycles (1916\,--\,2007). We train the CNN with the manually identified solar disk and plage. We first detect the solar limb and the North-South line in the suncharts, which enables the extraction of disk centre coordinates, radius, and P-angle. Following that, we use a CNN similar architecture to achieve accurate image segmentation for the identification of plages. We compare plage areas derived from the suncharts with those obtained from \ca\ full-disk observations, and find good agreement that demonstrates the potential application of such an ML technique for historical data. The results of this study further demonstrate the potential application of sunchart data to fill the existing data gaps in the KoSO multi-wavelength observations and contribute toward constructing a composite series over the last century.
\end{abstract}



\section{Introduction} \label{sec:intro}

\begin{figure*}[htb!]
 \centering
\includegraphics[width=\textwidth]{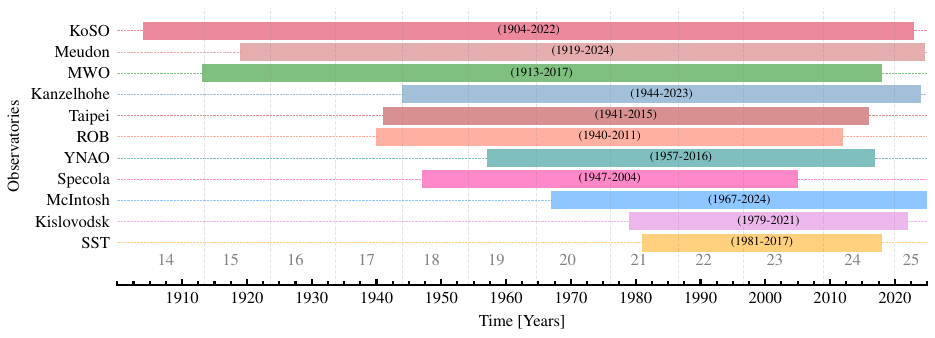}
\caption{Temporal coverage of solar drawings from various observatories, showing the duration of their observations in years. The KoSO provides the longest continuous dataset (1904\,--\,2022), followed by Meudon, MWO, Kanzelhöhe, and others.}
\label{sunchart_obs}
\end{figure*}

The study of solar magnetism has attracted significant interest over the past century, expanded by advancements in telescopic observations of prominent solar magnetic features such as sunspots (white light), plages (\ca), and filaments (\ha). Although naked-eye sunspot observations date back millennia, particularly within far eastern civilizations \citep{wittmann1987A&AS,yau1988QJRAS,hayakawa2017PASJ}, systematic telescopic observations only began in the early 17th century through the works of Harriot \citep{harriot1613}, Galileo \citep{galilei1613}, and Scheiner \citep{scheiner1615}, among others \citep{herr1978Sci,vaquero2009ASSL,arlt2020LRSP,carrasco2024ApJ}. Before photographic plates and digital imaging became available, detailed solar drawings provided crucial historical records of magnetic features, supporting reconstructions of solar activity extending back to the Maunder Minimum \citep{Hayakawa_2024}. Numerous archival datasets globally document solar features, notably from the Royal Observatory of Belgium (ROB, 1940\,--\,2011), Mount Wilson Observatory (MWO, 1913\,--\,2017), Specola Solare Ticinese (SST, 1981\,--\,2017), Meudon (1919\,--\,present), McIntosh archive (1967\,--\,present), and Kislovodsk solar station (1979\,--\,2021), all providing extensive coverage of solar magnetic activity \citep{priyadarshi2023ApJ}. Among these, the Kodaikanal Solar Observatory (KoSO) stands out due to its uniquely continuous record of annotated solar drawings (suncharts hereafter) spanning 1904\,--\,2022 \citep[119 years;][]{ravindra2020Ap&SS}. These KoSO suncharts are concurrently produced with multi-wavelength observations in white light \citep{mandal2017A&A,jha2022FrASS}, \ca\ \citep[393.37\,nm;][]{jha2024SoPh}, and \ha\ \citep[656.3\,nm;][]{chatterjee2017ApJ}, and contain detailed annotations of sunspots, plages, and filaments/prominences, offering an invaluable, comprehensive, and consistent dataset for long-term multi-wavelength studies of solar variability with its continuity and extensive temporal coverage (see \autoref{sunchart_obs}).

Plages in \ca\ images are valuable proxies for reconstructing pre-1970 solar magnetic fields due to the correlation between \ca\ intensity and magnetic field strength \citep{pevtsov2016A&A,mordvinov2020ApJ,shin2020ApJ,chatzistergos2019A&A, Yeates_2025}. They have been used to generate pseudo-magnetograms and investigate long-term solar variability \citep{virtanen2019A&A,virtanen2022A&A}, benefiting from their long lifetimes and structural stability \citep{zirin_1974}. Automatic and semi-automatic detection of plages \citep{priyal2014SoPh,chatterjee2016ApJ,barata2018A&C,chatzistergos2019A&A,chatzistergos2022FrASS,gharat2023RASTI} often suffer from false detections in archival data, as observed for KoSO \ca\ images \citep[post 1980;][]{jha2024SoPh,mishra2025ApJ}. These limitations motivate the use of machine learning (ML) methods, especially for complex datasets such as hand-drawn suncharts containing several kinds of markings apart from solar features. The generalization capability of ML architectures has greatly increased their reliance on the multiple problems of solar physics \citep[see][and references therein]{shin2020ApJ,upendran2020SpWea,raju2021SoPh,li2021RAA,deshmukh2023A&A, Chatterjee2022_natas, Chatterjee2023_apj}, with a comprehensive review provided by \citet{asensio2023LRSP}.

Recently, convolutional neural networks (CNN) and related ML algorithms have successfully been applied to detect filaments in \ha\ full‐disk images from Big Bear Solar Observatory \citep[BBSO;][]{zhu2019SoPh} and Global Oscillation Network Group and Kanzelhöhe Observatory \citep[GONG and KSO;][]{diercke2024A&A}, as well as to identify sunspots in Michelson Doppler Imager \citep[MDI;][]{chola2022GloTP}, Helioseismic and Magnetic Imager \citep[HMI;][]{chola2022GloTP,palladino2022}, Uccle Solar Equatorial Table \citep[USET;][]{sayez2023}.

Identification of solar features in suncharts is even more challenging due to the coexistence of diverse feature annotations and non-solar artifacts. Consequently, ML offers an optimal solution for accurate segmentation and classification in these complex datasets. Recently, \citet{priyadarshi2023ApJ} specifically applied an unsupervised ML technique called  ``K-means clustering" for filament detection using color space in digitized KoSO suncharts (1954\,--\,1976), digitised using a Canon EOS 800d camera. This work provided a proof-of-concept for the application of such techniques, which actually led to the complete digitisation of the suncharts at KoSO. However, neither this nor the prior studies have attempted to detect plages. Identifying plages using unsupervised ML was challenging in the past due to the indistinct color of plages, unlike filaments in KoSO suncharts, irregular morphology, and variability in sunchart quality and annotation styles. Additionally, the lack of well-labelled datasets has limited the use of supervised ML approaches. Overcoming these obstacles requires the creation of comprehensive, high-quality training sets and the design of transparent, reproducible CNN architectures that can accommodate the heterogeneous nature of archival solar images. 

Despite the advances in detection algorithms, the automatic identification of plages employing ML approaches has not yet been attempted. In this paper, we present a deep-learning-based supervised ML approach, specifically a CNN model, for the first automated identification of plages from the newly digitized KoSO suncharts covering the period from 1916 to 2007 (sunchart grid style is different for the pre-1916 period and plage marking unavailable after 2007). This automated plage detection will effectively address the data gaps in historical \ca\ full-disk images from KoSO, as highlighted in recent studies \citep{jha2024SoPh, mishra2024ApJ,mishra2025ApJ}, and will further aid the generation of pseudo-magnetograms through corroboration with available \ha\ dataset from KoSO.

The paper is structured as follows, in section \ref{Sec:data}, we describe the observational data utilized in this work; Section \ref{Sec:method} outlines the methodology employed for plage detection; in Section \ref{Sec:results}, we present our findings along with comparative analyses; and finally conclude in Section \ref{Sec:summary}.

\section{Data} \label{Sec:data}

KoSO has conducted solar observations across white light, \ca, and \ha\ wavelengths since 1904 using photographic plates and films. These have been digitized and calibrated for public access \citep{ravindra2013A&A,priyal2014SoPh,chatterjee2016ApJ,chatterjee2017ApJ,mandal2017A&A,chatzistergos2018A&A,jha2022FrASS,chatzistergos2023A&A}\footnote{\url{https://kso.iiap.res.in/data}}. The strength of this archive lies in its consistent optical setup and processing, though post-1980, it started to use films instead of photographic plates, and the image contrast also started to reduce.

\begin{figure*}[htbp!]
 \centering
\includegraphics[width=0.9\textwidth]{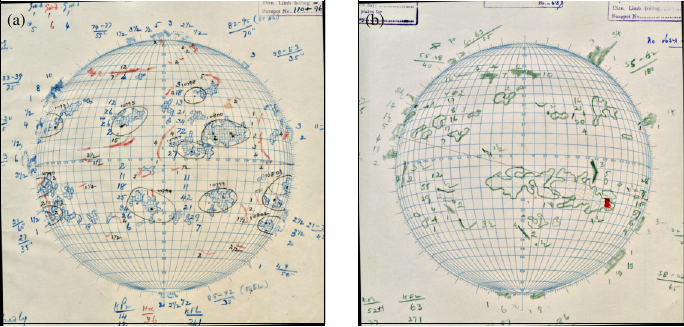}
\caption{Representative suncharts produced at KoSO. (a) A sunchart from the KoSO dated January 5, 1958, showing detailed hand-drawn annotations of sunspots, plages, filaments, and prominences observed by the KoSO. (b) Another sunchart produced at KoSO dated 1958 February 06, showing solar features with different color as the original observation was made by a different observatory. Please note that suncharts are contrast-enhanced for better visualisation.}
\label{sunchart_image}
\end{figure*}

Alongside photographic observations, systematic hand-drawn suncharts (\autoref{sunchart_image}a) have also been produced at KoSO. These suncharts document solar features such as sunspots, plages, filaments, and prominences by projecting the daily observations (white light for sunspots, \ca\ for plages and prominences, and \ha\ for filaments) onto a Stonyhurst latitude and longitude grid, which is divided into increments of 5$^{\circ}$ in both latitude and longitude directions. These drawings were typically created shortly after the plates or films were developed, minimizing the effects of scratches, fungal degradation, and other artifacts that can affect the long-term preservation of the photographic plates/films. The combined use of suncharts along with the digitized photographic plates and films offers a means to generate a consistent and reliable dataset of solar features spanning the period from 1904 to 2007.

Each solar feature is distinctly marked using specific colors: sunspots in black pencil, plages in blue ballpoint pen, filaments in red pencil, and prominences in blue pencil, as seen in \autoref{sunchart_image}(a). Additional observational metadata, including the position angle of the solar north (P-angle), heliographic latitude of the solar disk center (B-angle), and the observation time (white light data timestamp), is provided at the top of each chart. Typically, observers create one sunchart per day, based on the best available photographic observation. In instances where the daily observations from KoSO were unavailable, spectroheliograms from MWO or Meudon Observatory were used to supplement the records and prepare the suncharts. Particularly during the period of 1957\,--\,1958, an active exchange program of spectroheliograms was conducted between KoSO and a few other observatories under the scheme of the International Astronomical Union \citep{hale1931MWOAR,mausam1951,report1959MNRAS,hasan2010ASSP}. To clearly distinguish the observation taken from various observatories, different colors (for instance, green for plages and filaments) are used, as shown in \autoref{sunchart_image}(b). Further comprehensive details regarding these data are available in \citet{ravindra2020Ap&SS,priyadarshi2023ApJ}.

\subsection{Sunchart Digitization} \label{Sec:digitization}

The initial digitization of KoSO suncharts covered two solar cycles (cycles 19 and 20; 1954\,--\,1976) and was conducted using a Canon EOS 800D camera, as described in detail by \citet{priyadarshi2023ApJ}. More recently, however, the entire collection of KoSO suncharts spanning from 1904 to 2022 has been systematically digitized utilizing a high-resolution scanner capable of generating 6k $\times$ 6k images. For this purpose, at KoSO, the Zeutschel OS12002 scanner is used, which is specifically designed for high-resolution digitization of large-format materials such as books, drawings, and maps. It supports a maximum scanning resolution of 600 pixels per inch (PPI), enabling the accurate preservation of fine structural details. Two versions of these digitized suncharts have been produced: high-resolution (6k $\times$ 6k) and lower-resolution (1k $\times$ 1k) images, both stored in `.tif' format, offering distinct fields of view (cropped version). However, we used only high-resolution (6k $\times$ 6k) suncharts in this work. The filenames explicitly encode the presence or absence of specific solar features, enabling the quick identification of the sunchart contents without requiring direct inspection of the images. In this convention, the letter `k' denotes features annotated from KoSO and t or m from other observatories (MWO or Meudon), while `x' indicates the absence of a feature. For example, in ``sch$\_$19580105$\_$0754$\_$kkkk.tif'' all four features (sunspots, filaments, plages, and prominences) are drawn from KoSO observations, whereas in ``sch$\_$19580206$\_$0000$\_$xttt.tif'' the absence of sunspots is indicated by x, while filaments, plages, and prominences are included based on data from external observatories (MWO and Meudon) as shown in \autoref{sunchart_image}.

\begin{figure*}[htb!]
 \centering
\includegraphics[width=\textwidth]{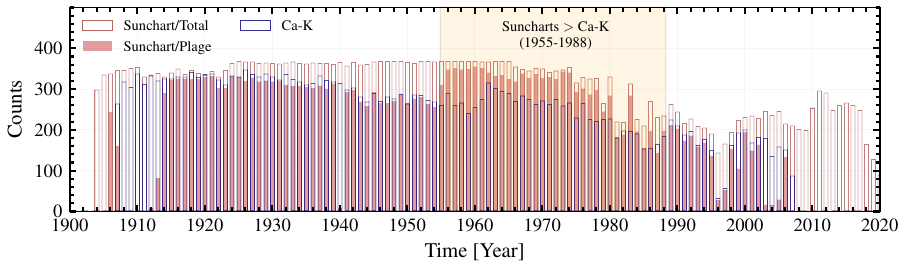}
\caption{Comparison of the annual number of \ca\ images with the total number of suncharts and those containing plage markings, represented by the number of observing days. This highlights periods where suncharts, particularly those with plages (1955\,--\,1988), can supplement or extend the \ca\ observational record.}
\label{sunchart_hist_obs}
\end{figure*}

The temporal distribution of these suncharts, measured in terms of observing days per year, is shown in \autoref{sunchart_hist_obs} as open red bars for the period 1904\,--\,2022. Furthermore, this histogram also compares the total number of suncharts with those specifically featuring plages (filled red bars), alongside the number of \ca\ observing days (open blue bars). The comparison highlights certain intervals (notably from 1955 to 1986) where the number of available suncharts exceeds the number of corresponding \ca\ images, emphasizing the utility of suncharts in filling existing observational gaps in the KoSO \ca\ dataset. Moreover, the suncharts are particularly valuable in addressing the challenge of accurately identifying plage regions after 1980, a period during which data degradation in photographic plates/films significantly affected the quality of \ca\ images in the digitized data.

In the present study, we utilize high-resolution (6k $\times$ 6k) digitized suncharts covering the period from 1909 to 2007 (31438 images) and  1916 to 2007 (29070 images) for the automated identification of disk and plages, respectively. Years prior to 1909 are excluded as many of the earlier suncharts do not include solar features overlaid on the Stonyhurst grid, as shown in Appendix (\autoref{sunchart_without_grid}(a)). We also did not utilize the data of the pre-1916 period due to the different pattern of the sunchart grid (blue-filled) for this period (Appendix \autoref{sunchart_without_grid}b). 

\section{Methodology}\label{Sec:method}
In this study, we employ a supervised ML model called U-Net to automate the identification of plages within KoSO hand-drawn suncharts. U-Nets are based on CNNs and are widely used as a highly effective tool, particularly in biomedical image segmentation tasks \citep{ronneberger2015,cicek2016arXiv}. In the domain of solar physics, recent studies have employed U-Net models successfully for the automatic identification of various solar phenomena, such as coronal jets \citep{liu2024ApJ} and solar filaments \citep{zhu2019SoPh}. Motivated by these promising results, we adopt the U-Net CNN architecture in our analysis to identify solar plages automatically on digitized suncharts from KoSO. 

The U-net architecture comprises fully convolutional layers structured in an encoder-decoder configuration, creating a distinctive U-shaped framework that demonstrates excellent performance even with limited training datasets. The encoder portion uses 3$\times$3 convolutions with ReLU nonlinearity and systematically reduces the spatial dimensions of the input images through max pooling. This results in hierarchical feature learning, which simultaneously extracts and preserves critical image features. After flattening the extracted features from the encoder into vectors \citep{raju2021SoPh}, the decoder section progressively restores the image to its original dimensions. It combines the encoded high-resolution features with those from up-convolutional operations, ultimately producing the segmented image as shown in \autoref{model}. The mathematical description of the U-Net model is presented in \autoref{appendix:Unet}.

We train U-nets to perform two segmentation tasks as described in the following two subsections. The training and inference tasks of our U-Net model are executed on an in-house Graphical Processing Unit (GPU) equipped with a Quadro RTX 8000 graphics card and 50 GB of VRAM.

\begin{figure*}[htb!]
 \centering
\includegraphics[width=18cm]{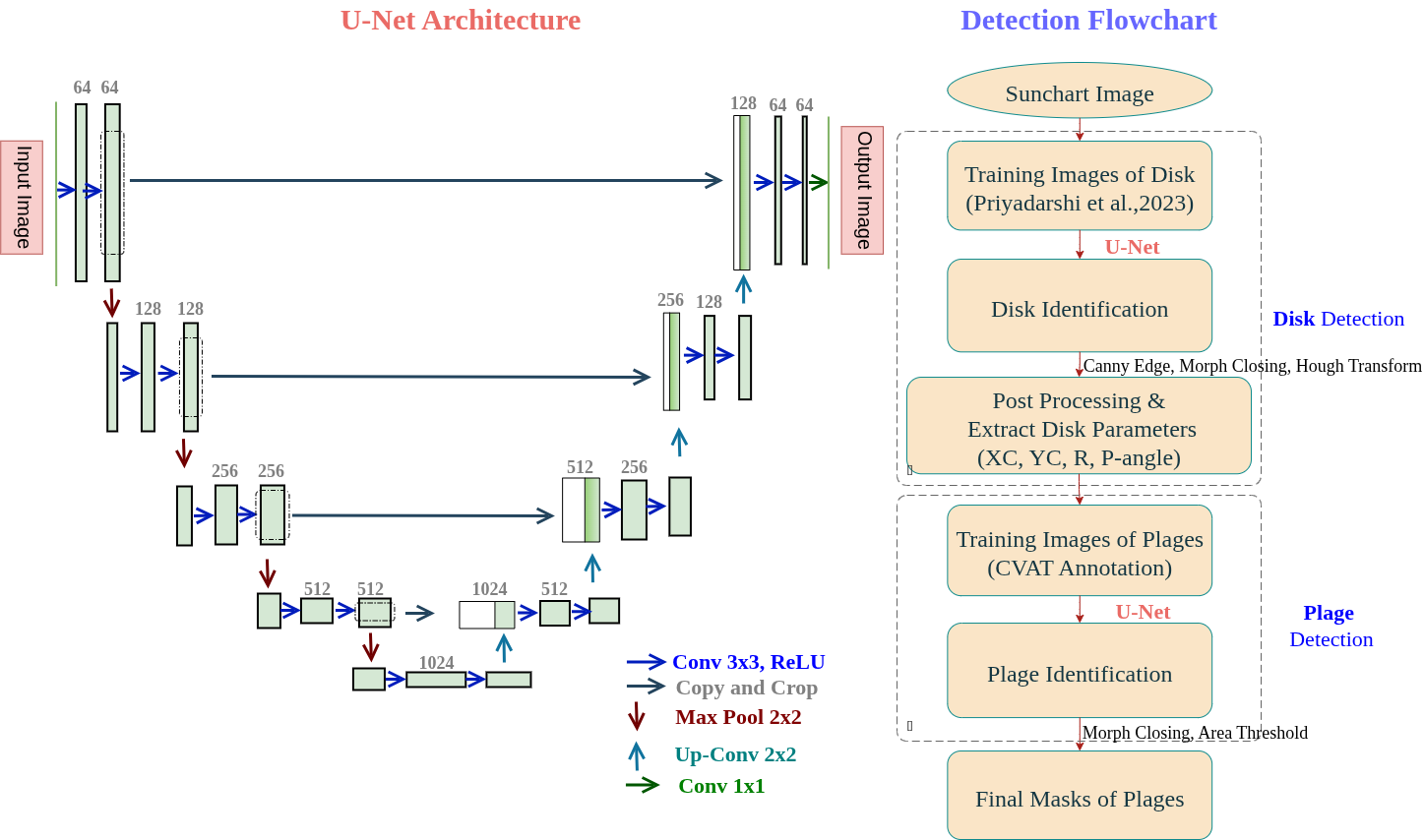}
\caption{(Left): U-Net architecture (adapted from \citet{ronneberger2015}) used for segmenting plages. The architecture consists of a contracting path (left) for feature extraction and an expansive path (right) for precise localization. Key components include 3$\times$3 convolutions with ReLU nonlinearity (blue arrows), 2$\times$2 max pooling (dark red arrows), 2$\times$2 up-convolutions (light blue arrows), and skip connections for combining high-resolution features from the contracting path with the upsampled output. (Right): Flowchart showing the steps for disk and plage detection in suncharts.}
\label{model}
\end{figure*}
\subsection{Disk Identification}\label{disk_idnt}

\begin{figure}[htb!]
 \centering
\includegraphics[width=8.5cm]{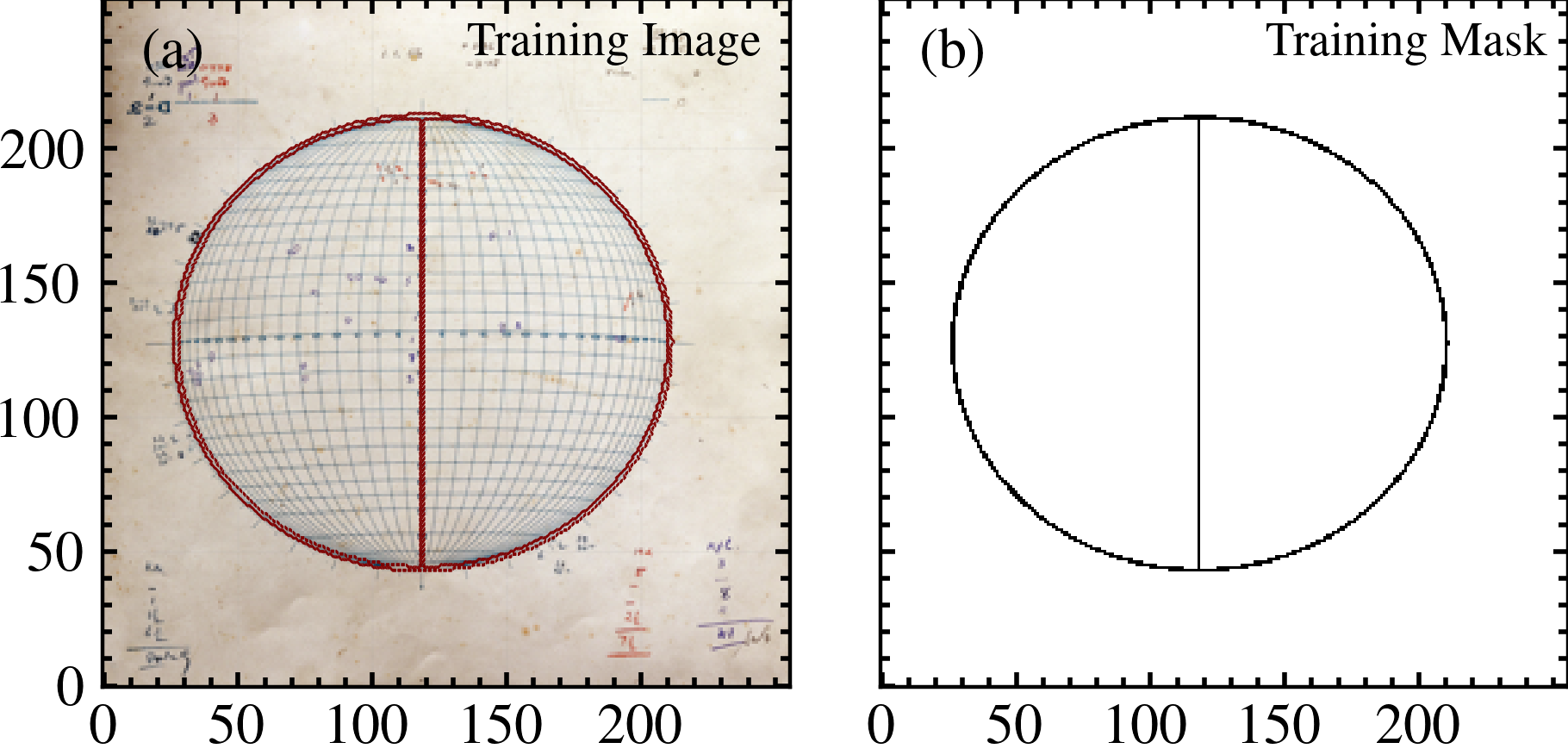}
\caption{(a) A representative training image with an annotated training mask (red). (b) The corresponding binary mask used for training the disk detection model.}
\label{aditya_train}
\end{figure}

Prior to the identification of plages, it is essential to accurately detect the solar limb and the North-South line in the suncharts to extract parameters such as disk center coordinates, radius, and P-angle, and assign physical coordinates to the plage locations. The extracted disk parameters are not only essential for plage identification but also serve as crucial information for the future detection of other solar features in suncharts, such as sunspots, filaments, and prominences. In previous work, \citet{priyadarshi2023ApJ} utilized the linear Hough transform technique for disk detection in old digitized suncharts covering the period 1954\,--\,1976 (solar cycles 19 and 20). However, employing the Hough transform across the entire dataset (1904\,--\,2022) proved inefficient due to significant variability in image quality over the extended observational timeframe. To address this challenge, we implemented a U-Net CNN model for robust disk detection across the complete data of digitized suncharts (1909\,--\,2007). For training the CNN model, we used previously digitized sunchart images from \citet[1954\,--\,1976;][]{priyadarshi2023ApJ}, along with corresponding binary masks representing the solar disk and central meridian line derived from known center coordinates, radii, and P-angle values extracted in the same study using Hough transform. From this subset, we systematically selected 7,310 high-quality sunchart images, which were subsequently resized to dimensions of 256$\times$256 pixels to standardize the input data for training the neural network. A representative example from the training dataset is presented in \autoref{aditya_train}(a and b). During the training procedure, pixel-wise differences between the predicted segmentation masks and the ground truth masks were quantified by the binary cross-entropy loss function defined as:
\begin{equation}
\mathrm{loss}(Y,\hat{Y}) = -\frac{1}{n^2}\sum_{i,j=0}^n \left[ y_{ij}\log\hat{y}_{ij} + (1-y_{ij})\log(1-\hat{y}_{ij}) \right]\,,
\label{eq1}
\end{equation}
where $Y = y_{ij}$ and $\hat{Y} = \hat{y}_{ij}$ are target and predicted segmentation maps correspondingly. We used the Adam optimizer \citep{kingma2017adammethodstochasticoptimization}, a first-order gradient-based adaptive optimization algorithm, for training the neural network \citep{kingma2017adammethodstochasticoptimization}. Adam dynamically adjusts learning rates during each training step, assigning individual adaptive rates for different parameters, thus improving convergence efficiency. The training process involved iterative passes over subsets (batches) of the dataset, where network parameters were updated after processing each batch according to the optimization strategy. An epoch is defined as one complete iteration through the entire dataset. Generally, larger batch sizes can enhance the performance of neural network classification or segmentation, although this choice must be balanced against batch size, model complexity, data dimensionality, and available computational resources (CPU/GPU memory). In this study, the batch size was set to 64, and the network was trained for a maximum of 25 epochs. The training was concluded at 25 epochs, as this point corresponded to the minimum training and validation loss, along with the Intersection over Union (IoU) score, indicating optimal model performance. With these settings, the training process took approximately thirty minutes to complete. 

\begin{figure}[htb!]
 \centering
\includegraphics[width=8.5cm]{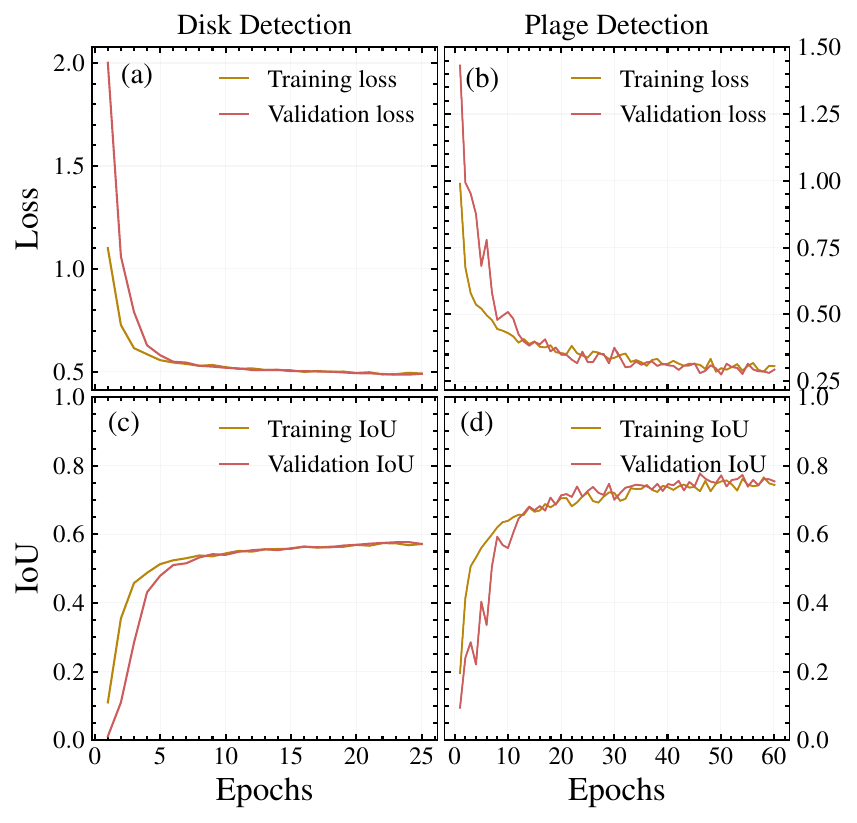}
\caption{Evolution of model loss and performance with epoch. (a) and (b) show the loss and Intersection over Union (IoU) curves, respectively, for the solar disk detection model. (c) and (d) show the corresponding loss and IoU curves for the plage detection model. Both models exhibit convergence with closely matching training and validation curves, indicating robust performance of the model.}
\label{iou}
\end{figure}

To evaluate the model's performance during training, we computed both the training and validation losses. The training loss quantifies the error between the predicted outputs and the true labels within the training dataset, while the validation loss measures the error on a separate validation dataset that is not included by the model during training. It is important to note that 24\% of the dataset is allocated for validation purposes in this work. The loss curves for both datasets are shown in \autoref{iou}(a). The convergence of these curves toward the end of training indicates that the model is neither underfitting nor overfitting, suggesting effective learning and generalization. In addition to the loss, we also evaluated the model using the  IoU metric for both training and validation datasets. IoU is widely used to assess the accuracy of object detection and image segmentation models. For a predicted region P and the ground truth region G, IoU is defined as:
\[
\text{IoU} = \frac{\text{Area of Overlap} \; (P \cap G)}{\text{Area of Union} \; (P \cup G)}.
\]
The IoU score ranges from 0 to 1, where 0 indicates no overlap and 1 indicates a perfect match between the predicted and ground truth regions. The IoU curves for the training and validation datasets are presented in \autoref{iou}(b). The consistent convergence of both curves further confirms that the model is learning effectively to detect the solar disk with reasonable accuracy.

\begin{figure*}[htb!]
 \centering
\includegraphics[width=\textwidth]{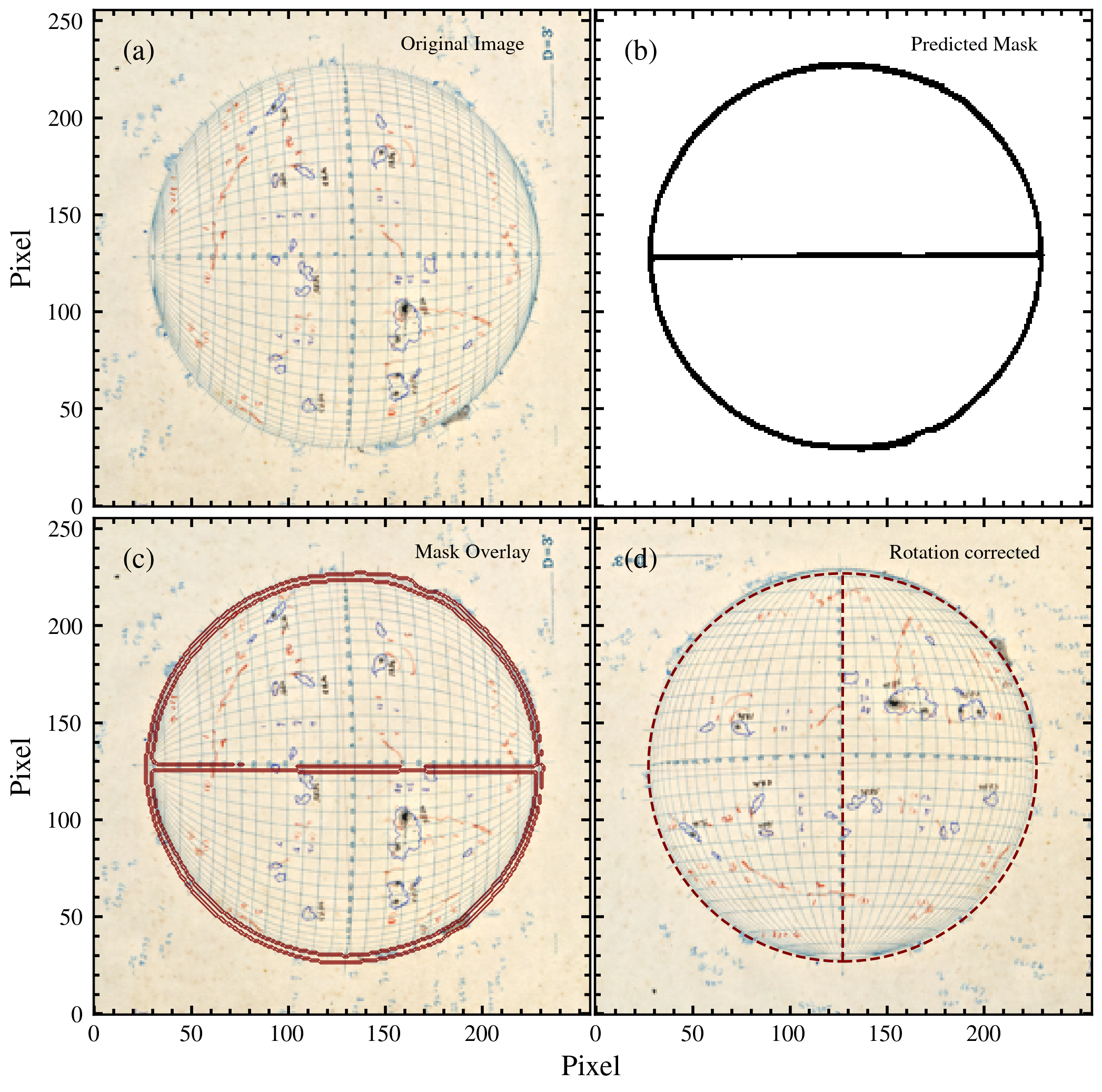}
\caption{(a) A representative image for a sunchart dated 1917 January 02. (b) Predicted disk mask generated by the trained model. (c) Overlay of the processed mask after post-processing using Canny edge detection and morphological operations (closing, erosion, dilation), with a contour on the original image. (d) Final rotation-corrected image with a circular fit (dashed red) applied using the Hough Circle Transform to align the solar disk with the vertical axis. Please note that suncharts are contrast-enhanced for better visualisation.}
\label{disk_identification}
\end{figure*}

After training, we use the model to infer the solar disk and north-south line from the digitized sunchart images resized to 256$\times$256 pixels, covering the period 1909--2007. An example of a predicted solar disk using the trained model is presented in \autoref{disk_identification}(b) for the sunchart shown in \autoref{disk_identification}(a). To refine the disk segmentation, we apply the Canny edge detector \citep{canny} and morphological operations such as closing, erosion, and dilation to the output of the model, as shown in \autoref{disk_identification}(c). The processed disk contour is then overlaid onto the original sunchart for the observation date of 1917 January 01 and is shown in \autoref{disk_identification}(c) using a red outline. This overlay highlights the accuracy and robustness of the model's prediction. Subsequently, we apply the linear and circular Hough transform methods \citep{hough} to extract geometric parameters of the solar disk. The linear Hough transform is used to determine the solar P-angle from the orientation of the detected central meridian line, while the circular Hough transform provides the center coordinates and radius of the disk. Specifically, the Hough transform identifies Hough circle peaks, in terms of the number of intersections, which correspond to the most probable circular features in the parameter space, representing the best-matching circles detected in the image. The resulting fitted circle and line are shown as red dashed contours over the rotation-corrected sunchart image in \autoref{disk_identification}(d), further validating the model's effectiveness. Since the model was trained and applied on downscaled 256$\times$256 images for computational efficiency, the extracted disk parameters (center, radius, and P-angle) were appropriately scaled to correspond to the original 6k$\times$6k resolution images for the full period from 1909 to 2007. We also manually corrected the P-angle for certain images (5\%) due to the poor quality of the corresponding sunchart data, which hindered accurate automatic estimation. This derived disk and P-angle information is subsequently utilized for automated plage detection, which is described in the following section. We also constructed a FITS standard metadata file for the sunchart by extracting the solar disk region, allowing the data to be easily integrated with SunPy \citep{sunpy_community2020}.\footnote{\url{https://docs.sunpy.org}} An example image generated using SunPy's Map object overlaid with the Stonyhurst heliographic grid (blue) shown in the Appendix (\autoref{sunchart_sunpymap}). The sunchart data, along with adequate metadata information, will be made publicly available in the future. This information will be necessary for any other studies that use these suncharts, e.g., for identifying other solar features, such as sunspots, filaments, and prominences, in suncharts.

\subsection{Identification of Plages} \label{plage_idnt}

\begin{figure}[htb!]
 \centering
\includegraphics[width=8.7cm]{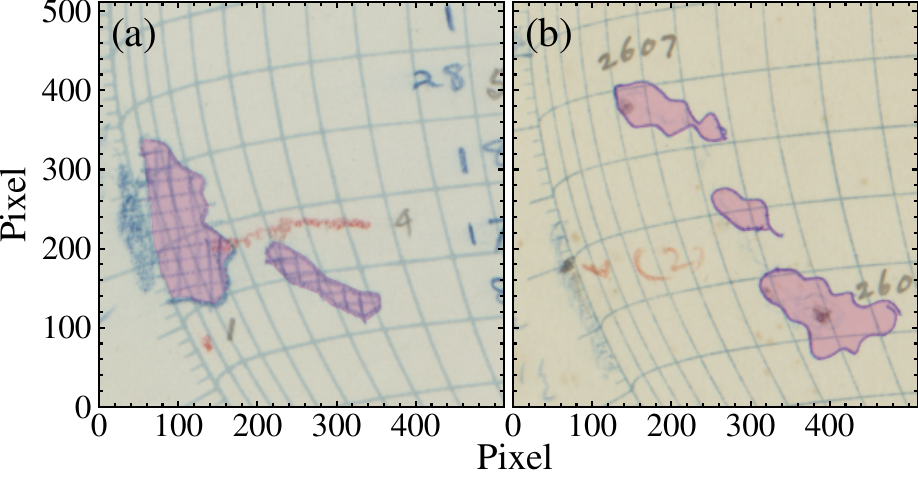}
\caption{Plage annotation using CVAT. (a) and (b) show input image patches from 1958 March 14 and 1917 January 28, respectively, with pink-filled contours indicating manually annotated plage region masks.}
\label{plage_training}
\end{figure}

The suncharts from KoSO have detailed records of various solar features, including sunspots, plages, filaments, and prominences, which are valuable for filling gaps in historical datasets. In the present study, we primarily focus on the detection of plages, along with accurate solar disk identification, to assess the potential applicability of ML techniques for the analysis of historical observations. For the identification of plages, we utilize the same U-net architecture previously used for solar disk detection, as described in \autoref{disk_idnt}. To prepare the training dataset, we first apply P-angle correction to the digitized sunchart images. Due to computational constraints, the original high-resolution 6k$\times$6k images are resized to a 2048$\times$2048 format. These resized images are then divided into 512$\times$512 patches using the `patchify' module to meet the input size requirements of the U-Net model during training. The corresponding ground truth masks for these image patches are generated through semi-automated annotation using the open-source Computer Vision Annotation Tool \citep[CVAT:][]{cvatai}\footnote{\url{https://app.cvat.ai/}}, which is widely used in the computer vision community. We utilize the AI-assisted annotation feature of CVAT, which leverages pre-trained machine learning models to support object detection, classification, and segmentation. This feature significantly improves annotation speed and accuracy by allowing users to quickly identify and mark regions of interest, thereby enhancing the overall efficiency of the training dataset preparation. For model training, we select Sunchart images from the period 1958 to 1968, as this interval captures many plage regions encompassing the maximum activity phases of two strong solar cycles (SC19 and SC20), for robust training. Additionally, we include images from 1917 to mitigate detection bias and to ensure the model learns to detect plages with diverse morphological characteristics, as plage structures from the earlier period differ notably from those in later drawings in terms of hatched lines. Examples of the training patches and their corresponding annotated masks are presented in \autoref{plage_training}, where the ground truth masks are also overlaid on the original images using red contours. A total of 7,875 training patches were utilized to train the model, with a random validation-to-training split ratio of 0.2.

Once the training patches are prepared, we prepare the U-net model where a ResNet-34 backbone \citep{nsiah2023} serves as the encoder, extracting low-dimensional features from the images. The encoder is pretrained on large-scale datasets, enabling transfer learning, which significantly reduces training time and enhances model performance, particularly in cases with limited training data. Following preprocessing, we apply data augmentation techniques to reduce the risk of overfitting and to help the model learn invariant features. This augmentation process introduces new transformations of the training images, including rotation, flipping, zooming, and brightness adjustments. For training, we utilize the Adam optimizer and the binary cross-entropy loss function, which is consistent with the configuration used in the disk identification model. However, we modified the training parameters, using a batch size of 32 and 60 epochs based on the optimal performance of the model with these hyperparameters. It took training the model for plage identification on a GPU, which required approximately 1.5 hours. The performance of the model is shown in \autoref{iou} (c and d), where both the training and validation loss curves show convergence after a certain number of epochs, indicating model stability and effective learning. The observed fluctuations in the loss and IoU curves, along with relatively lower IoU scores, are attributed to the morphological complexity of features in the suncharts, which poses a limitation for the model. Nonetheless, the model performs reasonably well, with the validation loss reaching 0.25 and IoU reaching a value of 0.8 at the end of 60 epochs, given the inherent variability and intricacy of the input data.

\begin{figure*}[htb!]
 \centering
\includegraphics[width=\textwidth]{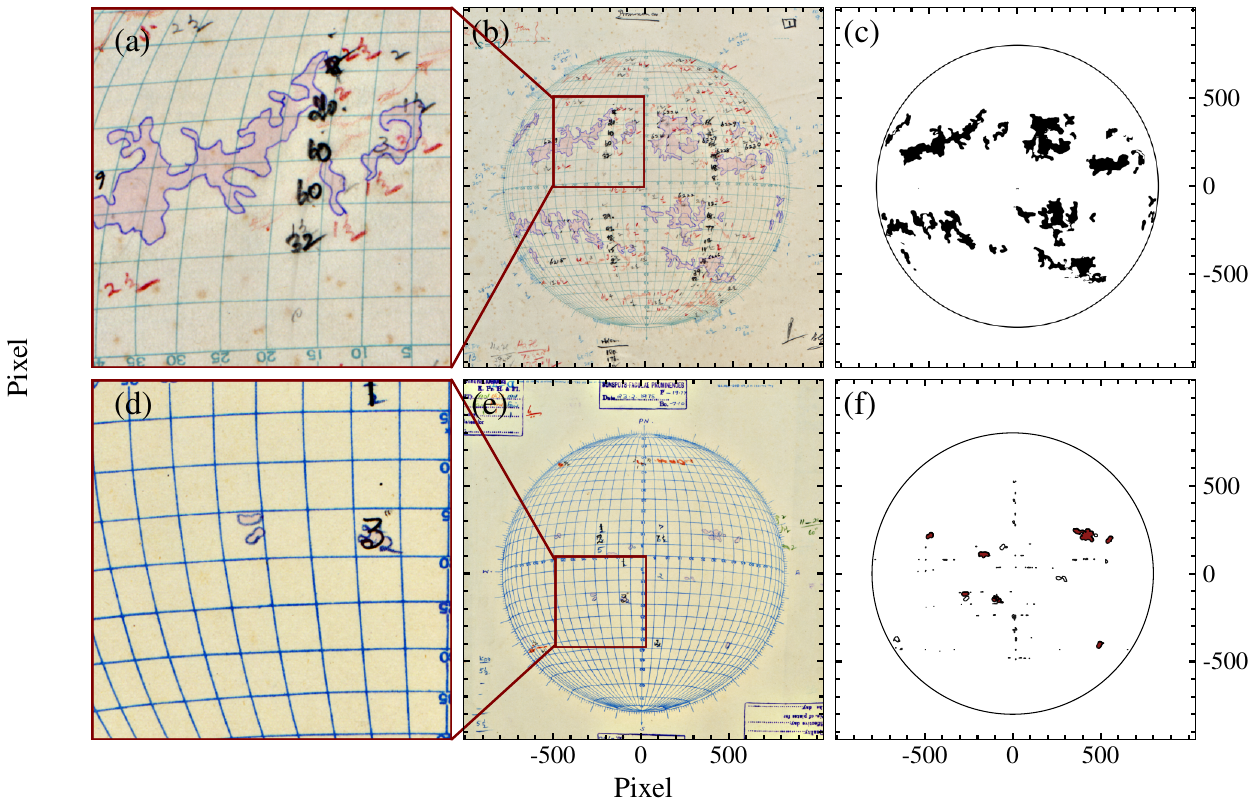}
\caption{Plage detection from KoSO suncharts. (a) A zoomed-in region of the KoSO sunchart observed on January 2, 1937, with the predicted plage contours shown in pink. (b) The full sunchart image from the same date, with predicted plages overlaid, and (c) the corresponding binary plage mask. (d) A zoomed-in patch from a sunchart with poor-quality annotations observed on 1975 February 23, and (e) the corresponding full-disk image with predicted plage regions. (f) Final processed binary mask from (e) after applying morphological closing and area-based filtering. Red-filled contours in (f) highlight the remaining plage regions after post-processing. Please note that Suncharts are contrast-enhanced for better visualisation.}
\label{plage_prediction}
\end{figure*}

After completing the training phase, the U-net model is applied to the full dataset (1909\,--\,2007) of digitized KoSO sunchart images to automatically identify plages. As for the training dataset, each sunchart is first corrected for its P-angle and then divided into 512$\times$512 pixel patches, which are used as input to the model for prediction. An example of the predicted patches for the sunchart observed on 1937 January 02 is shown in \autoref{plage_prediction}(a). The predicted mask patches are subsequently recombined to reconstruct the full-disk binary mask of detected plages. To ensure spatial consistency, regions outside the identified solar disk were masked out, eliminating false detections beyond the disk boundary. The resultant full-disk plage masks are shown in \autoref{plage_prediction}(b, c), with the detected plage locations overlaid in pink on the original sunchart image for visual validation. It is important to note that for the period spanning 1909 to 1915, the model was unable to detect plages due to the suncharts being blue-filled (\autoref{sunchart_without_grid}(b)) as mentioned in \autoref{Sec:data}, which differs from the annotation style used in the post-1915 charts. Consequently, the analysis of plages based on suncharts is initiated from the year 1916 onward.

These results demonstrate that the model generalizes well to data outside the training period, as shown in \autoref{plage_prediction}, which represents data from years not included in the training set (i.e., 1958\,--\,1968 and 1917). However, we face some challenges. The model occasionally misclassifies the sunchart grid lines as plages, and some plages are fragmented rather than detected as cohesive structures. To address these limitations, we apply morphological closing operations and area-based thresholding (A $>600$ pixel$^{2}$) to merge fragmented plage regions and suppress small false detections, such as grid lines or numbers written in suncharts. This post-processing step is shown in \autoref{plage_prediction}(d,e,f), using the prediction results from the sunchart dated February 23, 1975. The trained model and the processed binary plage masks for the entire period (1916--2007) will be made publicly available in the future.

\begin{figure*}[htb!]
 \centering
\includegraphics[width=\textwidth]{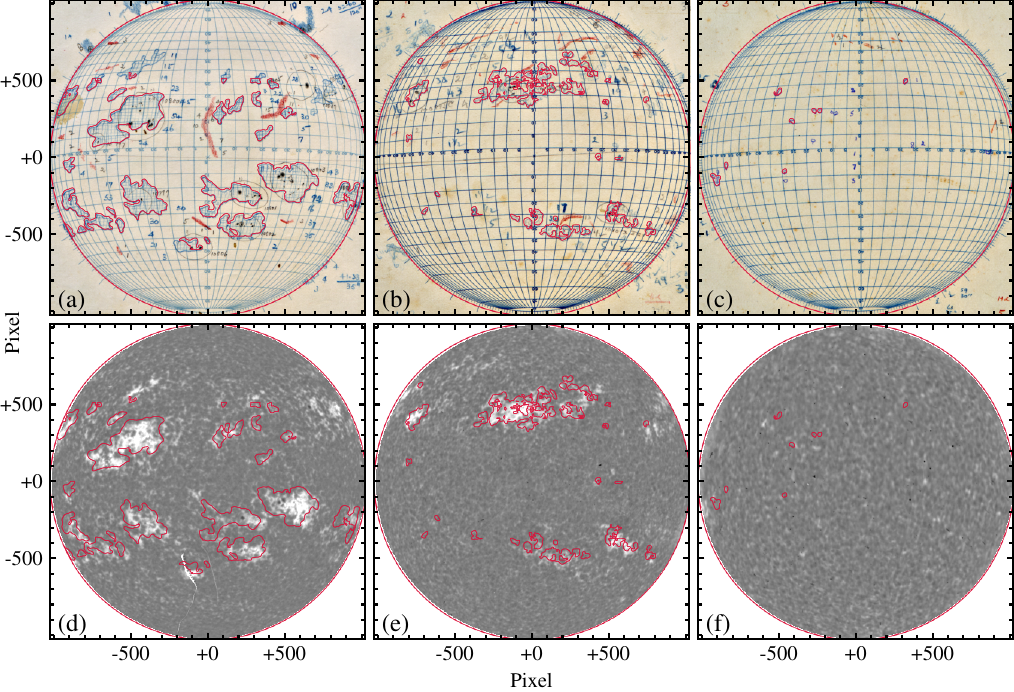}
\caption{Plage detection in different phases of the solar cycle. (a–c) Suncharts representing different phases of solar activity: solar maximum (1958 January 09), intermediate activity (1956 January 01), and solar minimum (1954 January 01). The predicted plage regions are overlaid as blue contours. (d–f) Corresponding \ca\ images from the same dates, aligned with the suncharts, and overlaid with the predicted plage masks from the respective suncharts (shown in blue). These comparisons demonstrate the model’s ability to identify plage structures across varying levels of solar activity. Please note that suncharts are contrast-enhanced for better visualisation.}
\label{activity_comp}
\end{figure*}

To evaluate model performance across varying solar activity conditions, we tested the predictions on suncharts from different phases of Solar Cycle 19, specifically the maximum (January 9, 1958), intermediate (January 1, 1956), and minimum (January 1, 1954) periods. The predicted plage masks are overlaid on the original sunchart images as red contours (\autoref{activity_comp} a–c), and compared with plage locations in co-temporal \ca\ full-disk images from KoSO (\autoref{activity_comp} d–f).

These comparisons indicate that the model performs robustly across different phases of the solar cycle. We note that some plages remain undetected during solar maximum, likely due to the increased morphological complexity and density of features. A schematic workflow summarizing the sequential steps involved in identifying the solar disk and plages from the sunCharts is presented in \autoref{model}.

\section{Results}\label{Sec:results}

\subsection{Plage Time-Latitude Diagram} \label{butterfly}

\begin{figure*}[htb!]
 \centering
\includegraphics[width=18cm]{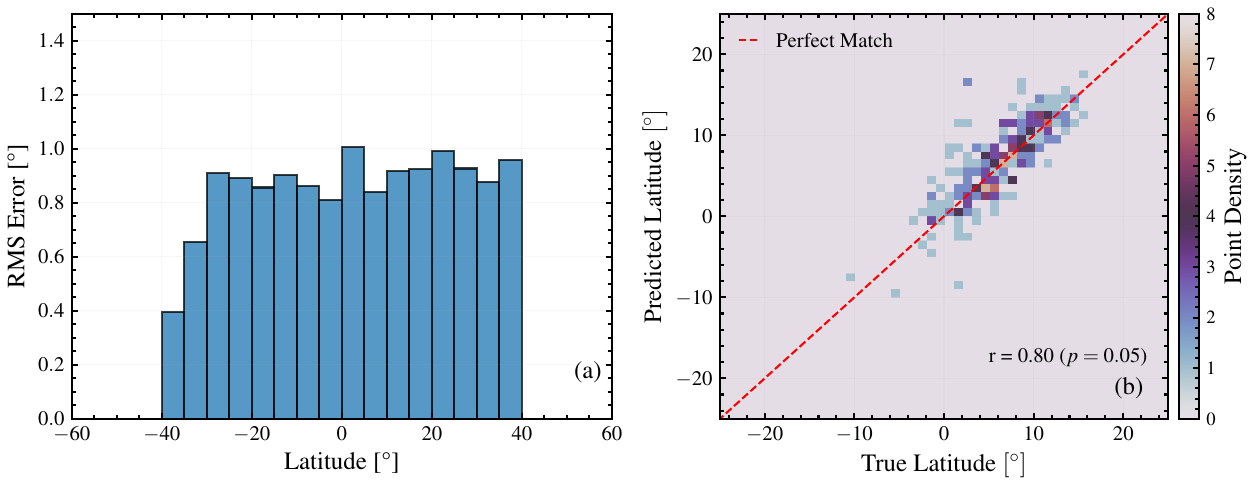}
\caption{Comparison of physical coordinates of detected plages with ground truth (a) Root-mean-square (RMS) latitude prediction error as a function of latitude. The model shows nearly uniform performance across all latitude bins.} (b) 2D histogram presentation comparing predicted vs. ground truth latitudes for plage centroids. The color scale indicates the point density of predicted and ground truth latitudes. The dashed line represents the one-to-one relation (perfect match), and the Pearson correlation coefficient (r = 0.80, p = 0.05) quantifies the level of agreement.
\label{butterfly_comp_gt}
\end{figure*}

Following the detection of plages on the suncharts spanning 1916\,--\,2007, binary masks of the identified regions are saved in a standardized 2048\,$\times$\,2048 format. To evaluate the model's performance in terms of extracting the spatio-temporal evolution of the Sun, we calculated the Root Mean Square (RMS) error in terms of latitude position between the ground truth and predicted plages. This error distribution is shown in \autoref{butterfly_comp_gt}(a) for the year 1959, where the spatial overlap between predicted and ground truth centroids is visualized in terms of error distribution. The correlation between the predicted and ground truth centroid latitudes is shown in \autoref{butterfly_comp_gt}(b), yielding a Pearson correlation coefficient (r) of 0.80. These results indicate a generally good agreement between the predicted and actual plage positions.  We have also calculated the average Balanced Accuracy (BA) score to be 0.94, defined as the mean of the True Positive Rate and the True Negative Rate. BA addresses the significant disparity in the number of pixels between the background (quiet Sun) and the foreground (plages), driven by the pixel-per-pixel success rate.

\begin{figure*}[htb!]
 \centering
\includegraphics[width=\textwidth]{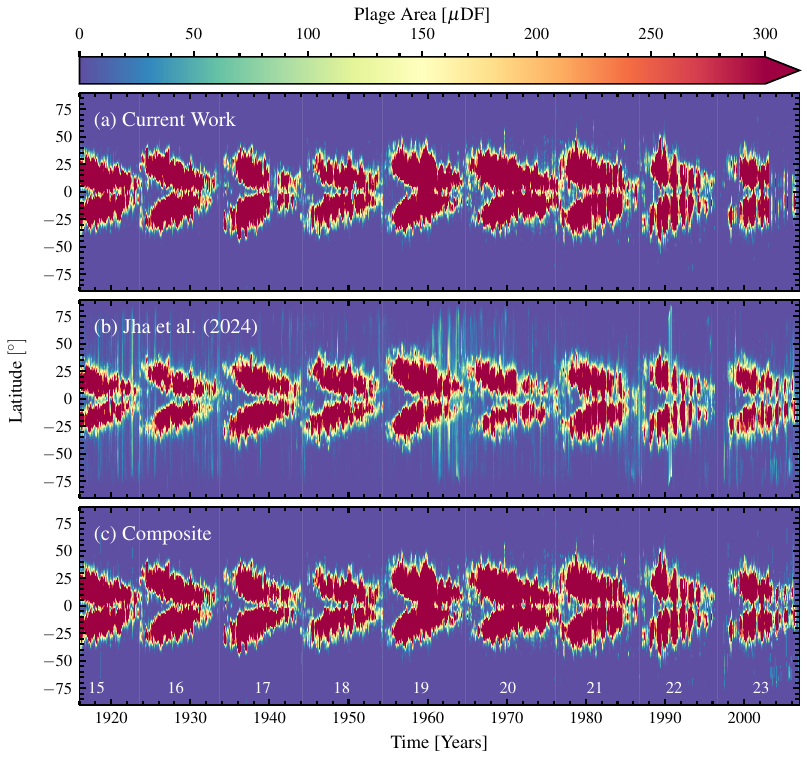}
\caption{Time–latitude distribution of plage area (in millionths of solar disk fraction, $\mu$DF) binned by 1$^{\circ}$ latitude for each observation day. In cases with multiple observations per day, values are averaged. (a) Results from the current study using the U-net-based detection on suncharts. (b) Corresponding butterfly diagram based on plage detections from \ca\ data \citep{jha2024SoPh}. (c) Composite series combining (a) and (b), using (b) to fill temporal data gaps in (a). The color bar represents plage area, highlighting latitudinal and solar cycle variation of plage area from 1916 to 2007 (solar cycles 15–23).}
\label{butterfly_comp}
\end{figure*}

The time-latitude distribution of plage areas, commonly referred to as the “butterfly diagram,” is presented in \autoref{butterfly_comp}. The diagram shown in \autoref{butterfly_comp}(a) is constructed using the plage masks derived from suncharts. The plage area was computed as the fractional disk area within 1$^\circ$ latitudinal bins across the range -90$^\circ$ to +90$^\circ$, similarly done in \citet{jha2024SoPh}. As with sunspots, plages tend to emerge at higher latitudes during the early phase of a solar cycle and progressively migrate toward the equator as the cycle advances, consistent with previous findings by \citet{chatterjee2016ApJ, priyal2017SoPh, jha2024SoPh}. For comparison, the butterfly diagram derived from the plage areas in \citet{jha2024SoPh} is shown in \autoref{butterfly_comp}(b). A qualitative comparison between \autoref{butterfly_comp}(a) and \autoref{butterfly_comp}(b) reveals that the sunchart-derived butterfly diagram exhibits reduced scatter and fewer spurious detections, particularly at higher latitudes. Notably, the artifacts present in the \ca\ plage masks, appearing as artificial latitudinal stripes around the years 1961 and 1990 due to artefacts resulting from degradation (or other reasons, such as scratches or fungi) in digitised \ca\ images, are absent in the sunchart-based masks. Although the sunchart record exhibits data gaps, particularly around 1940 and post-2003, most of the observational coverage remains intact. To address the remaining gaps, we constructed a composite time-latitude diagram by supplementing the missing sunchart-derived plage area data with corresponding values from the \ca\ plage masks. The resulting composite diagram, shown in \autoref{butterfly_comp}(c), provides a more continuous and complete temporal coverage. Minor missing data near the year 2007 remain and could be addressed in future work by incorporating additional \ca\ plage datasets from other archives.

In addition to the plage area–based butterfly diagram shown in \autoref{butterfly_comp}, we also compute the time-latitude distribution of plages using the centroid latitude of each plage region, following the approach adopted by \citet{chatterjee2016ApJ} and \citet{priyal2017SoPh}. The resulting distribution, derived from the sunchart-based plage masks, is shown in \autoref{plage_area_centroid}(a). For comparison, the corresponding diagrams from \citet{chatterjee2016ApJ} and \citet{priyal2017SoPh} are presented in panels (b) and (c) of \autoref{plage_area_centroid}, respectively. A clear improvement in temporal coverage is evident in the sunchart-derived dataset in comparison to \ca\ image-based plage masks \citep{chatterjee2016ApJ,priyal2017SoPh}. The gaps present in the \citet{chatterjee2016ApJ} and \citet{priyal2017SoPh} \ca\ datasets are largely attributable to the exclusion of lower-quality images during their processing, as discussed in \citet{jha2024SoPh}. Despite minor data gaps around 1940 and after 2003, the sunchart-based centroid latitude distribution offers a more exhaustive record of the long-term spatio-temporal evolution of plages.

\subsection{Plage Area Series} \label{plage_area}

\begin{figure*}[htb!]
 \centering
\includegraphics[width=\textwidth]{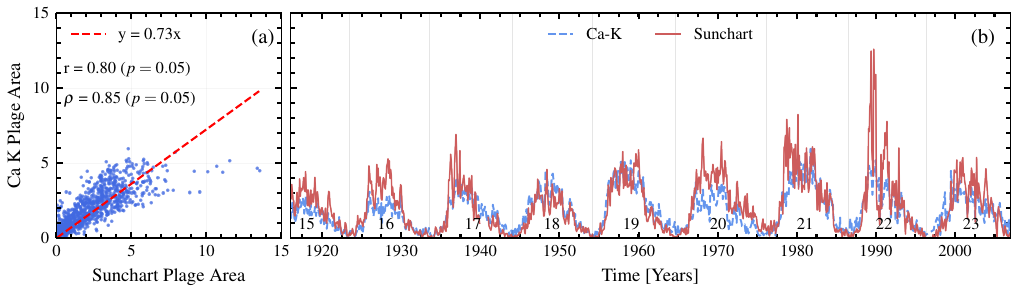}
\caption{(a) Scatter plot showing the correlation between monthly mean plage areas ($10^{4}\mu Hem$) derived from suncharts and \ca\ observations. The red dashed line represents a linear fit with a slope of 0.73. Pearson correlation coefficient ($r$=0.80) and Spearman rank correlation coefficient ($\rho$=0.85) both suggest a strong agreement between the two quantities (p=0.05). (b) Monthly averaged plage area series from Suncharts (in $10^{4}\mu Hem$) derived from sunchart-based detections (red) compared with corresponding values from \ca\ full-disk observations (blue) spanning solar cycles 15 to 23.}
\label{plage_area_series}
\end{figure*}

In recent times, there have been several studies that have investigated the long-term variation of plage area using \ca\ observations from various historical observatories. For instance, \citet{chatterjee2016ApJ} and \citet{priyal2017SoPh} analyzed KoSO \ca\ data, while \citet{chatzistergos2018A&A} utilized data from multiple archives, including Arcetri, Kodaikanal, McMath-Hulbert, Meudon, Mitaka, Mt. Wilson, Schauinsland, Wendelstein, and the modern Rome/PSPT series to reconstruct the long-term evolution of plage area. In this study, we present the comparison of the plage area derived from the KoSO sunchart-based plage masks spanning 1916–2007 with previous \ca\ plage area results, provided by (private communication with Theodosios Chatzistergos) \citet{theo2019SoPh,theo2020A&A,chatzistergos2022FrASS,jha2024SoPh} to verify our model output. A quantitative comparison between the monthly KoSO sunchart and \ca\ plage areas yields a Pearson correlation coefficient ($r$=0.80) and Spearman rank correlation coefficient ($\rho$=0.85), indicating a strong correspondence between the two independent datasets (see \autoref{plage_area_series}(a)). This shows the reliability of plage detection in suncharts from our model in a quantitative way.

The time series of monthly averaged plage area is shown in \autoref{plage_area_series}(b) as a red continuous line. For comparison purposes and further verification of the CNN model, we overlay the corresponding plage area obtained from KoSO \ca\ masks (blue dashed line). It is worth noting that in this \ca\ dataset, extremely poor-quality images that could not be processed were excluded. The two time series show a strong agreement overall, with noticeable deviations in solar cycles 16, 20, and especially 22. In cycle 22, a sudden spike is observed around 1989 in the sunchart series, attributable to significantly larger plage markings by the observer during that period, as evident in \autoref{sunchart_high_plage}(a). Additionally, a general trend of larger plage annotations is observed in suncharts after 1964, likely reflecting a shift in observer bias or the definition of plage boundaries. A notable drop in the sunchart plage area around 1990 arises from model failure in detecting plages due to their unusual representation in black ink for that year, see \autoref{sunchart_high_plage}(b). This mislabeling also resulted in physically implausible high-latitude detections. The data gap in the series near 1940 is due to the absence of plage markings in the suncharts. Notably, after 1978, the \ca\ plage area series (\autoref{plage_area_series}(b)) exhibits a decline in the number of available data points. This gap is supplemented by plage area measurements derived from suncharts. In future work, we plan to combine the sunchart-derived plage area with the \ca\ series to investigate the long-term variability of plage properties.

\section{Summary and Conclusion}\label{Sec:summary}

This study presents the first successful application of an ML approach for the detection of plages in historical suncharts spanning 9 solar cycles. Initially, we identify the solar disk and north-south line in suncharts over the period 1909\,--\,2007 using a U-Net model, extracting key disk parameters such as center coordinates, radius, and P-angle. These disk parameters will be useful for future studies in the analysis of other solar features (sunspots, filaments, prominences) in suncharts and will be available online in `fits' format. Subsequently, plages are segmented using a similar U-net architecture, and the resulting masks are saved in a standardized format for the period 1916\,--\,2007. We systematically derive physical coordinates of detected plages and compare them with those obtained from \ca\ full-disk observations to evaluate consistency. We construct time–latitude diagrams from the sunchart-derived plage masks and compare them against similar diagrams produced from \ca\ data, finding good agreement. To enhance temporal coverage, we also develop a composite butterfly diagram by integrating data from both sources. Furthermore, we compare the monthly variation in sunchart-derived plage area from 1916 to 2007 with that from \ca\ observations, resulting in a strong correlation ($r$=0.80 and $\rho$=0.85), confirming the effectiveness of the detection method. These results collectively demonstrate the efficacy and robustness of the proposed ML approach for automated plage detection in historical hand-drawn Suncharts. 
 
Despite the successful detection of plages in the KoSO suncharts, several limitations remain in the current model. The model fails to identify plages in the pre-1916 period and in 1990 due to inconsistencies in the representation of the Stonyhurst grid and variations in plage annotations in these datasets. This highlights the need to develop a more generalized and robust model capable of handling diverse sunchart formats across different observational periods. Additionally, suncharts from the pre-1909 period exhibit a significantly different grid structure, which has hindered the reliable detection of solar disks using the present method. Future work will focus on enhancing the detection method by incorporating training datasets from multiple solar cycles, thereby accounting for variations in plage appearance, particularly the color inconsistencies observed in years such as 1990. This will be achieved by standardizing training datasets using grayscale images to improve model generalization. Additionally, implementing a calibration procedure for plage area estimates will help mitigate observer-dependent biases inherent in the manual annotation of suncharts.
 
Given all the limitations of the current study, we believe the detected plage information from this century-long series of hand-drawn suncharts will make a valuable addition to the historical solar chromospheric observations. The \ca\-sunchart composite will not only enhance the observation statistics by mitigating data gaps but also help constrain the solar magnetic evolution in the past, complementing sunspot records.
While the current study has focused solely on the identification of plages, future extensions of this work will target the automated detection of other annotated solar features in suncharts, including sunspots, filaments, and \ca\ prominences. This would produce a more complete picture of the long-term spatio-temporal evolution of the solar magnetic field.


\begin{acknowledgments}

The data and suncharts used in this analysis were developed at the Kodaikanal Solar Observatory. We are grateful to the many observers who consistently carried out observations and prepared the suncharts, and to Mr. P Manikandan (Sathish) for their patient work in digitizing them. We gratefully acknowledge ARIES for providing the GPU server essential to developing our model. We gratefully acknowledge Theodosios Chatzistergos for providing the \ca\ plage area series and for his valuable suggestions. The funding support for DKM's research is from the Council of Scientific \& Industrial Research (CSIR), India, under file no.09/0948(11923)/2022-EMR-I. This research used version 7.0.1 \citep{stuart_j_mumford_2025_16638197} of the SunPy open source software package \citep{sunpy_community2020}. This study has utilized the SAO/NASA Astrophysics Data System's bibliographic services.

\end{acknowledgments}

\software{IDL, sunpy, tensorflow, keras}

\appendix

\section{U-Net Model Description}\label{appendix:Unet}

Here, we discuss the mathematical operations underlying the U-Net architecture. As mentioned in \autoref{Sec:method}, U-Net comprises an encoder–decoder pair. The encoder performs a sequence of convolutions with appropriate padding, followed by downsampling (e.g., max pooling), to extract hierarchical features while progressively reducing the spatial resolution. For encoder layer $Le$, the feature map $(O_{Le}^{en})$ is produced from the output $(O_{Le-1}^{en})$ of the previous layer according to \autoref{eq:app_eq1}, i.e., the convolution followed by a ReLU nonlinearity. At each level of the encoder and decoder, two consecutive convolutional operations are applied, each followed by a nonlinear activation function.

\begin{equation}
    O_{Le}^{en} = \text{ReLU}\left(\text{Conv}(O_{Le-1}^{en})\right)
    \label{eq:app_eq1}
\end{equation}

The central part of the ``U'' represents the region (``bottleneck layer") where the network encodes the most compact and abstract representation of the input features, effectively forming the bridge between the encoder and decoder. Within this bottleneck layer, both the feature dimensionality and spatial resolution are maintained to preserve consistency in feature representation. This compressed latent space is designed to retain only the most relevant information necessary for the accurate reconstruction of the input or the generation of the corresponding segmentation map. The decoder performs learnable upsampling to restore spatial resolution, utilizing transposed convolutions (also known as deconvolutions) with appropriate padding. These operations expand the feature maps back to the input scale. For the decoder layer $Ld$, the output feature map $O_{Ld}^{de}$ is computed from the preceding representation $O_{Ld-1}^{de}$ via a deconvolution-based upsampling, as specified in \autoref{eq:app_eq2}.

\begin{equation}
    O_{Ld}^{de} = \text{ReLU}\left(\text{DeConv}(O_{Ld-1}^{de})\right)
\label{eq:app_eq2}
\end{equation}

Skip connections constitute the defining mechanism of the U-Net, transmitting encoder feature maps to their decoder counterparts at matching resolutions. This fusion (typically via concatenation) preserves spatial detail and stabilizes gradient flow; the operation is shown in \autoref{eq:app_eq3}.

\begin{equation}
O_{Ld}^{de} = \text{concat}\left(O_{Ld}^{de}, O_{nL-Le}^{en}\right),
\label{eq:app_eq3}
\end{equation}
where $nL$ is the number of layers in the encoder. If the network has $nL$ levels, the output from the encoder layer $nL–Le$ has the same spatial dimensions as the decoder layer. Concatenating them aligns features pixel-by-pixel. In the U-Net, the output head is a 1×1 convolution that linearly projects the decoder features at each spatial location to the target number of classes, yielding the segmentation output ($O_{\text{final}}(x, y)$) as defined in \autoref{eq:app_eq4}. This operation preserves spatial resolution while performing channel-wise mixing only; see \citet{bastug2024} for further details.

\begin{equation}
    O_{\text{final}}(x, y) = \text{sigmoid}\left(\text{Conv}_{1 \times 1}\left(O_{\text{Ld}}^{de}\right)\right)
    \label{eq:app_eq4}
\end{equation}

\begin{figure*}[htb!]
 \centering
\includegraphics[width=8cm]{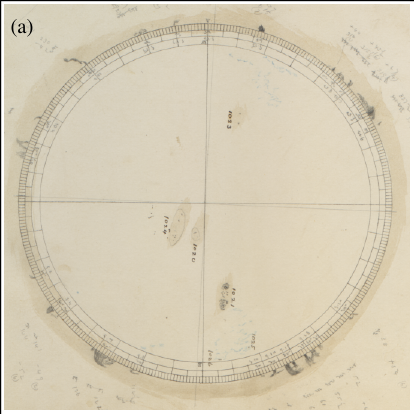}
\includegraphics[width=8cm]{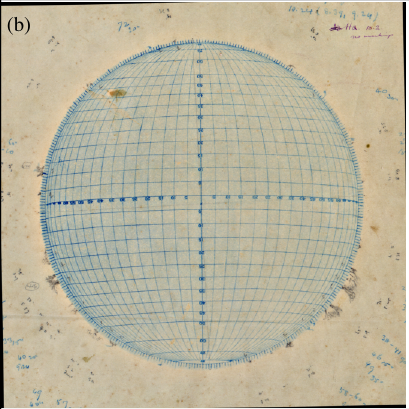}
\caption{(a): Sunchart observed on 1907 January 01, without the Stonyhurst grid overlay. (b): Sunchart observed on 1912 January 01, featuring a blue-filled Stonyhurst grid for coordinate reference and feature annotation. Please note that suncharts are contrast-enhanced for better visualisation.}
\label{sunchart_without_grid}
\end{figure*}

\begin{figure*}[htb!]
 \centering
\includegraphics[width=10cm]{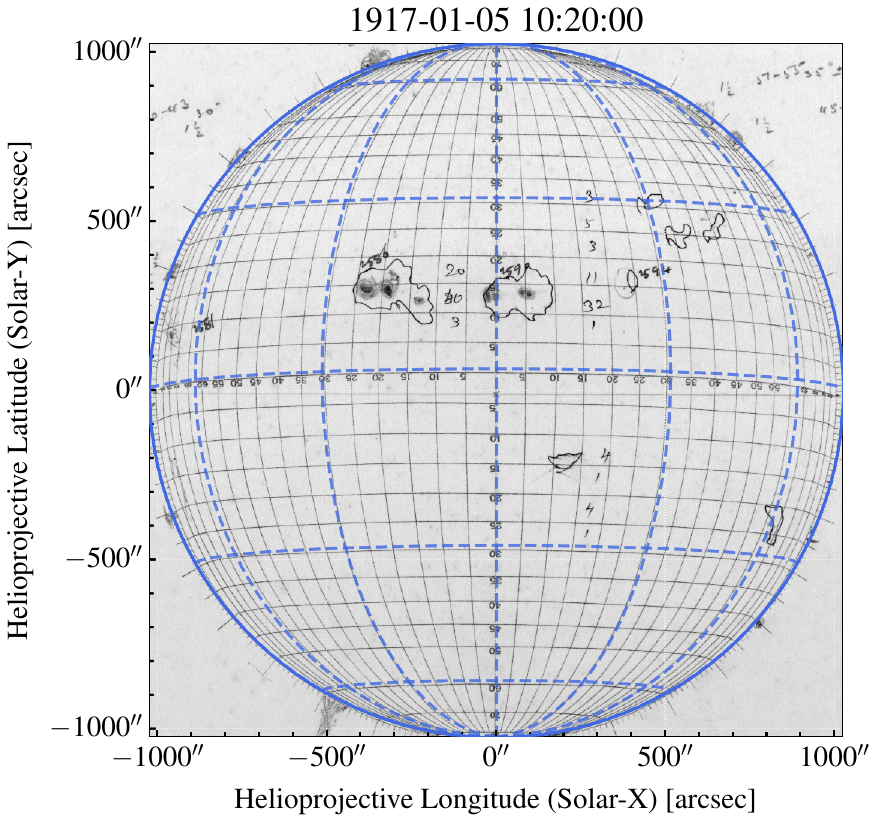}
\caption{Sunchart observed on 1917 January 05, shown as a SunPy map with the Stonyhurst heliographic grid (blue) overlaid.}
\label{sunchart_sunpymap}
\end{figure*}

\begin{figure*}[htb!]
 \centering
\includegraphics[width=\textwidth]{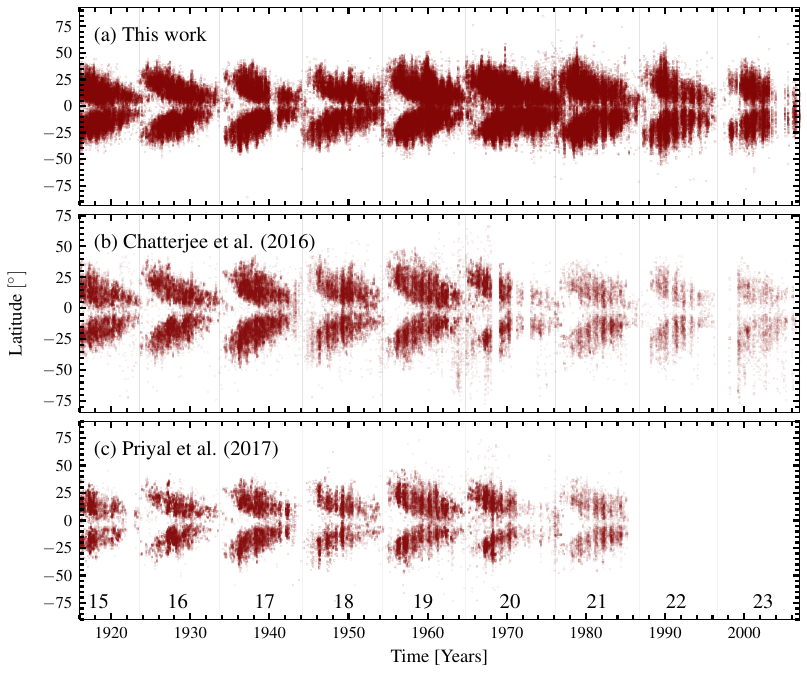}
\caption{Latitude-time distribution of the centroid positions of individual plage regions derived from (a) this work using CNN-based segmentation on suncharts, (b)  generated using plage regions from \citet{chatterjee2016ApJ}, and (c) \citet{priyal2017SoPh} on KoSO \ca\ images.}
\label{plage_area_centroid}
\end{figure*}

\begin{figure*}[htb!]
 \centering
\includegraphics[width=8cm]{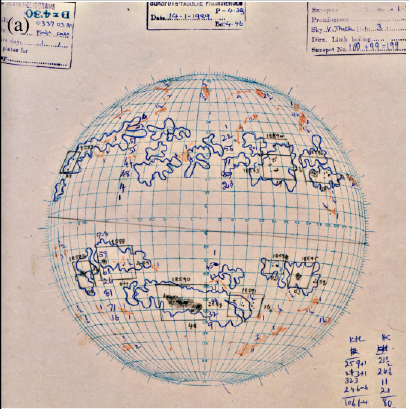}
\includegraphics[width=8cm]{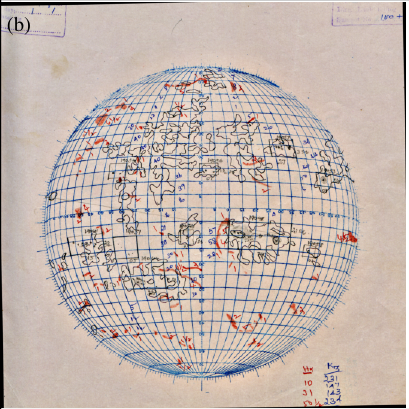}
\caption{(a): Sunchart observed on 1989 January 14, showing comparatively large-sized plage regions. (b): Sunchart from 1990 January 26, where plages are marked using different colours and are concentrated at relatively higher latitudes. Please note that suncharts are contrast-enhanced for better visualisation.}
\label{sunchart_high_plage}
\end{figure*}

\bibliography{sample7}{}
\bibliographystyle{aasjournal}



\end{document}